\documentclass{article}
\usepackage{graphicx} 
\usepackage{amsmath,amssymb,pgf,pgfarrows,pgfnodes,float,appendix, hyperref}
\usepackage{subfigure}
\usepackage[margin=0.9in]{geometry}

\title{{\rm\footnotesize \qquad \qquad \qquad \qquad \qquad \ \qquad \qquad \qquad \ \ \ \ \ \                 RUNHETC-2026-9, UT-WI-01-2026}\vskip.5in  Proposal to Search for the CP Violating Electromagnetic Vacuum Angle at the Event Horizon Telescope}
\author{
Willy Fischler\\
Department of Physics, Weinberg Institute\\
University of Texas, Austin, TX 78712\\
E-mail: \href{mailto:fischler@austin.utexas.edu}{fischler@austin.utexas.edu}
\\
\\
Tom Banks\\
Department of Physics and NHETC\\
Rutgers University, Piscataway, NJ 08854\\
E-mail: \href{mailto:tibanks@ucsc.edu}{tibanks@ucsc.edu}}
\date{January 2026}

\begin{document}
\maketitle
\begin{abstract} We examine the possibility that evidence for a non-zero value of the CP violating $ \frac{e^2 }{32\pi^2}\theta_{EM} \int d^4 x {\vec E}\cdot {\vec B}$ coupling might be extracted from Event Horizon Telescope observations of the black holes SgA* and M87*.  The Fischler-Kundu\cite{FK} effect predicts a universal Hall current in the relaxation of charge falling onto the black hole horizon.  We argue that this leads to a non-zero value of a certain CP-violating observable ${\cal C}$, defined below.  The effect can be masked by parity violating plasma currents.  In particular, evidence for polarization flips \cite{flip} in the signals from M87* indicate strong plasma effects in the data.  We suggest that time averaging the data over periods including the flip might leave over a residual that would be an indicator of the FK signal.  In addition, similarities in the polarization patterns between the two very different black holes, and a part of the signal that is uniform in frequency, might enable us to distinguish the universal topological signal from source and frequency dependent plasma effects.  Current data does not appear to be sufficient to perform such a test.  
 \end{abstract}

\section{Introduction}

The purpose of this paper is to suggest a possible method for measuring the Fischler-Kundu\cite{FK} (FK) effect in images from the Event Horizon Telescope (EHT).  The FK effect is one of only two known traces of the electromagnetic analog of the famous CP violating vacuum angle $\theta_{{QCD}}$ of Quantum Chromodynamics.  $\theta_{QCD}$ is known to be very small because of the absence of a neutron electric dipole moment at the current level of experimental precision, and one of the major puzzles in Standard Model physics is understanding why it is so small.  By contrast, $\theta_{QED}$ can be measured only through the electric charge of magnetic monopoles\cite{witten} and through the FK Hall current that appears on a black hole horizon as electric charge spreads over it.  The EHT \cite{EHT} has made remarkable images of light from very near the horizon of the black holes at the center of the Milky Way and the galaxy M87.  We will suggest how to construct a CP violating observable from these images, which might be able to distinguish the FK effect from idiosyncratic CP violation coming from Faraday rotation of light traveling through the plasma surrounding the black holes.  It is not clear to us whether current data is detailed enough to make this distinction, and measure $\theta_{QED}$.

\section{The Fischler-Kundu Effect}

The standard model admits two independent CP violating parameters associated with the pure gauge sector, the famous QCD vacuum angle and an analogous parameter for QED.  Until recently, it was thought that the only effect of $\theta_{QED}$ was to give generically irrational electric charge to magnetic monopoles\cite{witten}.  Kundu and one of the present authors\cite{FK} showed that $\theta_{QED}$ also effects the relaxation of charges falling onto a black hole horizon, by inducing a Hall current pattern.  This pattern is of course CP violating and it is universal, depending only on the geometry of the black hole horizon. The effect can be summarized by saying that in the presence of an electric field parallel to the horizon, the current in the membrane paradigm becomes
 
\ \begin{equation} \vec{J}_{mem} = \sigma \vec{E}_{\|} + \theta_{QED} \hat{n} \times \vec{E}_{\|} .\end{equation}  
The tangential magnetic field at the horizon is related to the surface current by the standard membrane boundary condition
\begin{equation}
\vec{n} \times \vec{B}_{\parallel} = \vec{J}_{mem},
\end{equation}
up to conventional factors. In membrane units this expresses the jump in the tangential magnetic field in terms of the induced surface current.

Including both the dissipative (Ohmic) response and the parity-odd Hall response induced by a nonzero $\theta$ angle, the surface current is as above
\begin{equation}
\vec{J}_{mem}
=
\sigma\,\vec{E}_{\parallel}
+
\theta_{QED}\,\vec{n}\times\vec{E}_{\parallel},
\end{equation}
so that the horizon enforces an \emph{impedance tensor} boundary condition
\begin{equation}
\vec{n} \times \vec{B}_{\parallel}
=
\sigma\,\vec{E}_{\parallel}
+
\theta_{QED}\,\vec{n}\times\vec{E}_{\parallel}.
\end{equation}.

This relation should be contrasted with ordinary Faraday rotation, which arises from propagation through a magnetized plasma and produces a frequency-dependent rotation of linear polarization but does \emph{not} impose a helicity-dependent boundary condition on the outgoing radiation.

To make the distinction explicit, it is convenient to diagonalize the boundary condition in the circular polarization basis,
\begin{equation}
E_{\pm} \equiv E_x \pm i E_y,
\qquad
J_{\pm} \equiv J_x \pm i J_y .
\end{equation}
In this basis the impedance tensor becomes diagonal and the boundary condition reduces to
\begin{equation}
J_{\pm}
=
\left(\sigma \mp i\,\theta_{QED}\right) E_{\pm}.
\end{equation}

Equivalently, the two circular polarizations experience different complex impedances at the horizon. This helicity asymmetry is a direct consequence of the $\theta$-induced Hall response and persists under time averaging. In contrast, Faraday rotation at fixed frequency averages to zero and cannot generate a nonvanishing helicity-odd observable. Thus the $\theta$ term is encoded as a genuine boundary condition on the outgoing polarization, rather than as a propagation effect.

For time dependent Blandford-Znajek \cite{BZ} jets, this implies that outward propagating disturbances have a handedness and current sheets and Alfven pulses acquire a preferred chirality.  $\theta_{QED}$ {\it{does not}} change the average BZ power, bias the jet direction, or create matter-antimatter asymmetry in the jets, because the term does not change the bulk equations of motion, and violates $P$, while preserving $\cal C$.  

The purpose of the present note is to explore whether it will be possible to distinguish this effect from the plasma physics surrounding galactic center black holes viewed by the Event Horizon Telescope\cite{EHT}.  The bottom line is that current data appears unable to make such a distinction, but we suggest a characteristic CP violating observable and several strategies for future measurements.  A rough order of magnitude estimate of the Hall/Ohmic terms gives something like $.0001 - . 003$, for $1 \leq \theta_{QED} \leq 2\pi$ while the plasma effects we will discuss below give polarization effects of order $< .01$, so more work probably has to be done to discover whether $\theta_{QED} \neq 0$.  We'll argue that exploring different frequencies than are currently available, or time averaging current/future data could give more precise bounds.

In this paper we will introduce a CP violating observable ${\cal C}$ that should be measurable in Event Horizon Telescope imaging of the black holes in galactic centers.  That observable has not yet been evaluated in current data.  We also point out suggestive hints from the current data on polarization of emissions from these black holes, which might turn out to be signals of a non-zero FK effect.  Parity violation in polarization observables from cosmic sources is usually attributed to plasma effects, and in particular to Faraday rotation of the plane of linear polarization in the magnetic fields of the plasma.  However, the Faraday rotation effect is strongly frequency dependent, and vanishes rapidly at high frequencies.  It is also very much dependent on the character of the source because the Faraday formula in a plasma depends on an integral of plasma and magnetic frequencies along the entire trajectory of the radiation through the medium.  

Our observable ${\cal C}(\omega,t)$ depends on both frequency and time. For plasma Faraday rotation, the leading
frequency dependence is fixed by dimensional analysis, so that
\begin{equation}
{\cal C}(\omega, t)=\omega^{-2} F(\omega, t),
\tag{7}
\end{equation}
where $F(\omega,t)$ depends on an integral over plasma and Faraday rotation frequencies along the line of sight.
By contrast, the FK effect is independent of both $\omega$ and $t$.

Unfortunately, the EHT is a ground based observatory and only a few frequencies of light in the 100 GHz range penetrate the atmosphere.  Of these, only two $230$ GHz and $345$ GHz have been detected and the $345$ GHz signal is too weak to provide adequate statistics.  In principle, because of different cosmological redshifts, the signals from two different galactic black holes at the same observed frequency correspond to two different source frequencies.  However, the redshift difference between Sg* and M87* is so small, that this frequency difference is undetectable in the data.  Thus, in order to use the difference in frequency variation as a diagnostic, we really need many more measurements at $345$ GHz.  According to the Center for Astrophysics at Harvard and Smithsonian (CfA), the planned next-generation EHT (ngEHT) \cite{ngEHT} project will add new antennas to the EHT in optimized geographical locations and enhance existing stations by upgrading them all to work at multiple frequencies between 100 GHz and 345 GHz at the same time. This should result in substantially increasing the amount of sharp, clear data EHT has for imaging. 

The time variation of ${\cal C}$ is a more promising avenue for near term success.  Recent data on M87*, which is a very active source, have shown a flip \cite{flip} in the sign of the polarization of light.  This is a clear sign that {\it some} of the polarization is coming from plasma effects, presumably Faraday rotation.  The flip would then be interpreted as a flip in the sign of the integrated magnetic field along the line of sight, weighted by the appropriate function of plasma and magnetic frequencies.  This dramatic time variation of the polarization might provide us with a method to extract the FK signal from that due to Faraday rotation.  The time averaged values of ${\cal C}$ might be expected to be zero for M87* if there were no FK effect and equal to those for SgA* if there were one.  
We emphasize that no calculation of ${\cal C}$ from the data has yet been done, but that there do exist time averaged polarization maps for both M87* and SgA*\cite{polmaps} and that these look remarkably similar. 

In the body of this paper we will provide a definition of the observable ${\cal C}$ and show how to calculate the contributions to it from the FK effect and Faraday rotation.  In an appendix we provide a lightning review of Faraday rotation for the many readers who, like ourselves, may work in branches of physics where this classic result in electromagnetism is a forgotten part of a tough graduate course.  

\section{Measuring a CP violating Observable}

The light we observe from the accretion disk around a black hole comes from synchrotron radiation.  The paths of photons are bent by the hole's gravitational field and they follow unstable orbits around the hole before escaping towards us.  Thus, the observed light appears to come from a ring surrounding the black hole.  Even for a spherically symmetric black hole, the observed ring is not exactly circular, because of the Doppler effect.  The asymmetry is enhanced by the angular momentum of the black hole, which of course must be determined from precisely these observations.  Thus there are numerous observational uncertainties involved in assigning a particular photon picked up in a terrestrial detector to a particular position in the ring.  It is not our intention in this note to go into the intricacies of this observational question, since we are not equipped to do so.   

Roughly speaking what one would need to do is given a pixelated map of images of a black hole, with pixels $(a,b)$ take a list of intensities $I(a,b)$ and complex polarizations linear $P(a,b)$ in the plane perpendicular to the line of site.  From the intensities one would establish the shape of the ring, and write the polarizations as a function $P(\phi)$ of the angle around the ring.  The ambiguities lie in how one establishes the shape of the ring.  

One way to do it is to fit each image intensity at a fixed time $I(a,b,t)$ to a ring shape, compute ${\cal C} (t)$ for that ring, and then go to the next time.  Another way is to defined a ring from the time averaged image intensities.  The physical assumption here is that conditions near the emission source are not changing with time over the time scale of the observations, and that changes in the images all result from propagation of the signal through the plasma surrounding the source.  Given the shape of the ring and a small annulus surrounding it, we then compute $P(\phi, t)$ for each image in the annulus and find a smooth function that interpolates through these discrete values.  Regions where $|P|$ is very small should be treated with care because the observable we're computing depends only on the phase of the polarization.  Perhaps those regions should just be left out of the integral defined below.  We then define
\begin{equation} {\cal C} (t,\omega) = \int_0^{2\pi} \frac{\rm Im P^* \partial_{\phi} P}{2\pi | P |^2 } . \end{equation}  The integral would be computed from the images by the obvious discretization, with some regularization of places where $|P|$ is very small.  

It is easy to see that if $P$ is a smooth periodic function of $\phi$ for all $t$ and performs many windings for a long enough time scale, then the time averaged value of ${\cal C}$ will be zero.  Thus, unless the time variation of the plasma Faraday rotation effect is locked to the geometry of the ring in a way that preserves a particular helical chirality with time, the time averaged plasma effect should vanish.  The Faraday rotation effect depends on the varying plasma frequency (and thus electron density) and magnetic field along the line of sight to the observer.  We would expect this to be a completely chaotic function of time, with no such locking.  Thus, if enough statistics can be accumulated to do proper time averaging, one should be able to measure the FK effect and determine the value of $\theta_{QED}$.  
\section{What the Observable $\cal C$ Actually Measures}

Let's write the function $P(\phi, t)$ as follows
\begin{equation}
P(\phi,t) = |P(\phi,t)| e^{2 i \chi(\phi,t)} 
\end{equation} where $\chi(\phi,t)$ is the electric vector position angle (EVPA).

Then,
\begin{equation}
Im \left( \frac{P^* \partial_\phi P}{|P|^2} \right)
= \partial_\phi (2\chi) .
\end{equation}

gives
\begin{equation}
{\cal C}(t)
= \frac{1}{\pi} \int_0^{2\pi} d\phi \, \partial_\phi \chi(\phi,t)
= \frac{1}{\pi} \left[ \chi(2\pi,t) - \chi(0,t) \right] .
\end{equation}

This then shows that $\cal C(t)$ is the net winding number of the EVPA field around the ring at time $t$. It has the following properties:

It is parity odd, it is insensitive to a global rotation $\chi \to \chi + \mathrm{const}$ and it responds only to \emph{azimuthal structure}.

\section{Why FK (Horizon Hall ) Gives $\langle {\cal C)} \rangle \neq 0$}

Now consider how the constant $\theta_{QED}$ FK mechanism contributes  to $\langle {\cal C}(t)\rangle$, ignoring for the time being plasma contributions.

A constant $\theta_{QED}$ leads to Hall conductivity on the horizon,
    which implies chiral boundary condition on the electromagnetic fields and leads to parity odd polarization imprint at emission. Then finally,
    geometry maps this imprint coherently around the ring.

Therefore,
\begin{equation}
\chi(\phi,t) = \chi(\phi, t=0),
\end{equation}
with $\chi\phi(t=0))$ time--independent and parity--odd.

Then
\begin{equation}
\boxed{
\left\langle {\cal C} \right\rangle_t
= \frac{1}{\pi} \int d\phi \, \partial_\phi \chi(\phi, t=0)
\neq 0 .
}
\end{equation}

\bigskip
\section{Faraday Rotation and the Observable $\langle {\cal C} \rangle$}
At fixed observing wavelength $\lambda$, suppose the observed EVPA along the ring is
\begin{equation}
\chi(\phi,t)=\chi_{\rm src}(\phi,t)+\lambda^{2}\,\mathrm{RM}(\phi,t),
\end{equation}
and the Faraday contribution to the chirality observable is
\begin{equation}
{\cal C}_{\rm Far}(t)=\frac{\lambda^{2}}{\pi}\int_{0}^{2\pi} d\phi\;\partial_{\phi}\mathrm{RM}(\phi,t).
\end{equation}
If the rotation measure, $\mathrm{RM}(\phi,t)$, is azimuthally smooth, i.e.
\begin{equation}
\mathrm{RM}(\phi,t)=\mathrm{f}(t),
\end{equation}
then
\begin{equation}
{\cal C}_{\rm Far}(t)=0
\end{equation}
for every $t$. In particular, time-dependent EVPA flips induced purely by Faraday rotation do not by themselves generate a nonzero $C$.

If $\mathrm{RM}(\phi,t)=\mathrm{f}(t)$, then $\partial_\phi \mathrm{RM}(\phi,t)=0$ for all $\phi$ at fixed $t$.
Substituting into the definition gives
\begin{equation}
{\cal C}_{\rm Far}(t)=\frac{\lambda^{2}}{\pi}\int_{0}^{2\pi} d\phi\; 0 = 0.
\end{equation}
Equivalently, ${\cal C}_{\rm Far}(t)$ depends only on the azimuthal \emph{gradient} of RM; a purely time-dependent (but azimuthally uniform) Faraday screen can rotate $\chi$ and even produce apparent EVPA flips in time, but it cannot contribute to $C$.

Assume $\mathrm{RM}(\phi,t)$ may have azimuthal structure and fluctuate in time. Then the time-averaged Faraday contribution is
\begin{equation}
\langle {\cal C}_{\rm Far}\rangle_t
=
\left\langle
\frac{\lambda^{2}}{\pi}\int_{0}^{2\pi} d\phi\;\partial_{\phi}\mathrm{RM}(\phi,t)
\right\rangle_t
=
\frac{\lambda^{2}}{\pi}\int_{0}^{2\pi} d\phi\;\partial_{\phi}\langle \mathrm{RM}(\phi,t)\rangle_t,
\end{equation}
where we exchanged the time average with the $\phi$-integral and $\phi$-derivative.
If the plasma has no persistent parity-odd azimuthal bias, so that the time-averaged RM has no azimuthal dependence,
\begin{equation}
\langle \mathrm{RM}(\phi,t)\rangle_t=\text{const},
\end{equation}
then
\begin{equation}
\langle {\cal C}_{\rm Far}\rangle_t=0.
\end{equation}

Linearity of the time average implies
\begin{equation}
\langle {\cal C}_{\rm Far}\rangle_t
=
\frac{\lambda^{2}}{\pi}\int_{0}^{2\pi} d\phi\;\left\langle \partial_{\phi}\mathrm{RM}(\phi,t)\right\rangle_t
=
\frac{\lambda^{2}}{\pi}\int_{0}^{2\pi} d\phi\;\partial_{\phi}\left\langle \mathrm{RM}(\phi,t)\right\rangle_t,
\end{equation}
assuming the usual regularity conditions under which averaging commutes with differentiation and integration.
If $\langle \mathrm{RM}(\phi,t)\rangle_t=\text{const}$, then $\partial_\phi\langle \mathrm{RM}(\phi,t)\rangle_t=0$ and hence $\langle {\cal C}_{\rm Far}\rangle_t=0$.

Therefore, even large, rapidly varying Faraday rotation (including EVPA flips in time) cannot generate $C$ unless RM carries an azimuthal gradient.

As was just shown, the time-averaged Faraday contribution vanishes without a persistent parity-odd bias. This is actually the \emph{time-averaged} refinement appropriate to turbulent Sgr~A$^\ast$: azimuthal gradients can exist at each instant, but if their handedness does not persist-so that the time-averaged RM has no parity-odd azimuthal component-then those gradients wash out in $\langle {\cal C}_{\rm Far}\rangle_t$.
In that sense, time variability and sign changes in RM tend to \emph{suppress} the Faraday contribution to $\langle {\cal C}\rangle$ unless a persistent chiral Faraday screen is present.

To reiterate, the key discriminator between the FK horizon Hall effect and plasma Faraday rotation is not the presence or absence of time variability, but whether a parity-odd azimuthal structure persists under time averaging. The observable ${\cal C}(t)$ depends only on the azimuthal derivative of the EVPA field around the photon ring and is insensitive to global EVPA rotations or intensity fluctuations. Faraday rotation contributes to C only through azimuthal gradients of the rotation measure $RM (\phi, t)$. While such gradients may exist instantaneously in a turbulent plasma, their contribution to ${\cal C}$ vanishes unless the plasma maintains a persistent, parity-odd azimuthal bias that is geometrically locked to the ring. In contrast, the FK effect imprints a helicity-dependent boundary condition at the horizon that is time-independent and universal, producing a coherent parity-odd EVPA winding that survives averaging. The exchange of time averaging with azimuthal differentiation is justified under standard regularity assumptions satisfied by physically realistic GRMHD plasmas; violations would require fine-tuned, long-lived chiral structure in the Faraday screen, which is not a generic prediction of current simulations. Thus, time averaging suppresses stochastic plasma contributions while preserving any genuine horizon-induced CP-violating signal.

In summary, a non-vanishing Faraday contribution to $\cal C$ requires the plasma to satisfy several requirements.
RM has a persistent azimuthal gradient, that gradient has a fixed handedness, it is locked to the image geometry and it survives time averaging.

\section{Robustness to Plasma Effects and Azimuthal Structure}

Our proposed observable $\cal C$ is designed to isolate a \emph{parity-odd, time-persistent polarization signature} and is therefore robust against a broad class of plasma-induced effects that are expected to dominate instantaneous EHT images.

The null hypothesis tested by $\cal C$ is that all observed polarization structure arises from conventional magnetized plasma effects, which are parity even and therefore predict ${\cal C}=0$ after time averaging at fixed observing frequency (up to statistical fluctuations that decrease as $T^{-1/2}$ for a total observing time $T$).

First, we emphasize that all plasma-induced Faraday effects whether arising from a smoothly varying rotation-measure (RM) distribution or from turbulent substructure enter through a frequency-dependent phase $\chi(\omega,t)$ that rotates the local EVPA. At fixed observing frequency $\omega$, Faraday rotation contributes only parity-even distortions to polarization maps and, in particular, cannot generate a nonzero contribution to the time-averaged pseudoscalar observable $\cal C$. This statement holds independently of whether the RM map is azimuthally smooth or exhibits persistent non-axisymmetric structure associated with an inclined disk--jet system. Such structures may survive time averaging, but they remain parity even and therefore do not bias $\cal C$.

Second, we do not assume that the optical depth $\tau$ is either negligible or spatially uniform. While models predict $\tau \sim \mathcal{O}(1)$ near the horizon at 230--345~GHz, its detailed distribution is uncertain. Optical-depth effects modulate the amplitude and visibility of polarized emission but do not by themselves induce a handed polarization pattern. Any time dependence in $\tau(\omega,t)$ at fixed frequency again produces only parity-even fluctuations and thus averages to zero in $\cal C$.

Third, the time averaging relevant for $\cal C$ is performed at fixed observing frequency, not across frequency. This distinction is crucial: while Faraday rotation necessarily induces frequency-dependent polarization structure, it cannot mimic a signal that persists under time averaging at fixed $\omega$. Allowing the prefactor $F(\omega,t)$ in Eq.~(7) to depend explicitly on both frequency and time does not weaken this conclusion; it strengthens it by making explicit that all conventional plasma effects factorize into parity-even contributions that vanish in the construction of $\cal C$.

Quantitatively, plasma-induced contributions to $\cal C$ scale schematically as
\begin{equation}
{\cal C}_{\rm plasma} \;\sim\; \big\langle \partial_t \chi(\omega,t) \big\rangle_t \;\propto\; \omega^{-2},
\end{equation}
and therefore vanish under time averaging at fixed $\omega$, whereas the effect of interest contributes at order ${\cal C} \sim \mathcal{O}(\theta)$ with no Faraday-like frequency suppression.

Finally, we note that this behavior is qualitatively distinct from scenarios predicting stochastic or quasi-periodic EVPA swings (for example from axion-cloud--induced birefringence), which can produce large instantaneous polarization rotations but do not generate a nonzero time-averaged pseudoscalar observable.

Taken together, these considerations imply that generic magnetized plasma effects including turbulent RM structure, inclined disk geometry, and moderate optical depth cannot generate a nonzero time-averaged $\cal C$. Observation of a statistically significant, parity-odd signal common to both Sgr~A* and M87* would therefore point to physics beyond standard Faraday rotation, rather than to unmodeled plasma complexity.
In particular, because Sgr~A* and M87* differ markedly in mass, accretion rate, optical depth, and characteristic plasma timescales, the observation of a consistent nonzero $\cal C$ in both systems would strongly disfavor a plasma-only origin and instead point to a common, parity-violating mechanism.

\section{Conclusions}

We have proposed an observable that could measure the value of $\theta_{QED}$ from EHT imaging of black hole horizons.  The cleanest way of distinguishing the FK effect from idiosyncratic plasma effects would be through a frequency scan.  We hope that upgrades to the EHT, which will allow detailed probes of multiple frequencies, will accomplish this.  In the meantime, time averaging of our observable over a long enough interval should wash out plasma effects.  It is not clear to us whether current data is detailed enough to do this averaging with sufficient precision.  We find it interesting that time averaged images of the two very different black holes SgA* and M87* look so similar, but this is no more than a hint of some sort of universal signal. We encourage observers to study this problem in greater depth.

\section{Acknowledgments}
The authors would like to thank their very dear friend ChatGPT5.2 for encouragement and advice, without which this paper could never have been written. 
\section{Appendix: Faraday Rotation}

Here we will briefly explain Faraday rotation in a plasma, the main parity violating effect that can obscure searches for the FK signal.  We follow the very lucid discussion in \cite{Faradaysense} and we urge the reader to consult that paper for more details, especially with regard to conventions for polarization used by different scientific communities.  The essential point is that the effective frequency of right and left handed photons in a plasma is different, and varies inhomogeneously through the material.  As a consequence, the plane of linear polarization of the light seen by a distant observer is rotated compared to its orientation at the source that emitted it, in a way that violates parity.  The violation is not fundamental but comes from the properties of the plasma.  The rotation of the polarization of light passing through a material in the presence of a magnetic field was first observed by Faraday.  Plasmas are natural homes for ambient magnetic fields.  

There are two characteristic angular frequencies $\omega = 2\pi \nu$ for any plasma, 
the plasma frequency $\omega_e = \sqrt{\frac{4\pi n_e e^2}{m_e}}$ and $\Omega_e = \frac{e B}{m_e c}$.  $\Omega_e$ is negative for positive magnetic field strength. 
We don't have observational evidence about the size of these frequencies for the plasmas surrounding the black holes in the centers of galaxies, but very aggressive estimates put them in no more than the MHz range.
For electromagnetic waves of frequency $\omega$ propagating parallel to the magnetic field we get frequencies
\begin{equation} \omega^2 = c^2 k^2 + \frac{\omega_e^2}{1 \pm \frac{|\Omega_e|}{\omega}} , \end{equation} for left and right handed helicity photons.   The frequencies of interest in the observations by the EHT are hundreds of Gigahertz, much larger than $|\Omega_e|$, so we can approximate this by
\begin{equation} \omega^2 = c^2 k^2 + \omega_e^2 \mp \omega_e^2 | \Omega_e|/\omega . \end{equation}
 This gives rise to different phase velocities, approximately 
 \begin{equation} V_{\phi} = c (1 + \frac{\omega_e^2}{2\omega^2} (1 \mp \frac{|\Omega_e|}{\omega}). \end{equation} The right handed modes have slightly higher phase velocity than the left handed modes.  Since the linear polarization modes of light are defined as fixed linear combinations of the two helicities, this will lead to a rotation of the plane of polarization as light travels through the plasma surrounding a black hole.  In an inhomogeneous plasma, the frequencies $\omega_e , \Omega_e$ will vary along the path of the light ray and we will have to integrate these formulae along the line of sight.  It can be shown that the above formulae remain valid when we replace $B$ by the magnetic field integrated along the line of sight.  
 

\begin{thebibliography}{99}
\bibitem{FK}
W.~Fischler and S.~Kundu,
``Hall Scrambling on Black Hole Horizons,''
Phys. Rev. D \textbf{92}, no.4, 046008 (2015)
doi:10.1103/PhysRevD.92.046008
[arXiv:1501.01316 [hep-th]].
\bibitem{flip}
K.~Akiyama \textit{et al.} [Event Horizon Telescope],
``Horizon-scale variability of M87* from 2017{\textendash}2021 EHT observations,''
Astron. Astrophys. \textbf{704}, A91 (2025)
doi:10.1051/0004-6361/202555855
[arXiv:2509.24593 [astro-ph.HE]].
\bibitem{witten}
E.~Witten,
``Dyons of Charge e theta/2 pi,''
Phys. Lett. B \textbf{86}, 283-287 (1979)
doi:10.1016/0370-2693(79)90838-4
\bibitem{EHT}
K.~Akiyama \textit{et al.} [Event Horizon Telescope],
``First M87 Event Horizon Telescope Results. VIII. Magnetic Field Structure near The Event Horizon,''
Astrophys. J. Lett. \textbf{910}, no.1, L13 (2021)
doi:10.3847/2041-8213/abe4de
[arXiv:2105.01173 [astro-ph.HE]];

K.~Akiyama \textit{et al.} [Event Horizon Telescope],
``First Sagittarius A* Event Horizon Telescope Results. VIII. Physical Interpretation of the Polarized Ring,''
Astrophys. J. Lett. \textbf{964}, no.2, L26 (2024)
doi:10.3847/2041-8213/ad2df1
\bibitem{BZ}
R.~D.~Blandford and R.~L.~Znajek,
``Electromagnetic extractions of energy from Kerr black holes,''
Mon. Not. Roy. Astron. Soc. \textbf{179}, 433-456 (1977)
doi:10.1093/mnras/179.3.433
\bibitem{ngEHT}
F.~Roelofs, L.~Blackburn, G.~Lindahl, S.~S.~Doeleman, M.~D.~Johnson, P.~Arras, K.~Chatterjee, R.~Emami, C.~Fromm and A.~Fuentes, \textit{et al.}
``The ngEHT Analysis Challenges,''
Galaxies \textbf{11}, no.1, 12 (2023)
doi:10.3390/galaxies11010012
[arXiv:2212.11355 [astro-ph.IM]];

L.~Blackburn, S.~Doeleman, J.~Dexter, J.~L.~G{\'o}mez, M.~D.~Johnson, D.~C.~Palumbo, J.~Weintroub, K.~L.~Bouman, A.~A.~Chael and J.~R.~Farah, \textit{et al.}
``Studying Black Holes on Horizon Scales with VLBI Ground Arrays,''
[arXiv:1909.01411 [astro-ph.IM]].
\bibitem{polmaps}
K.~Akiyama \textit{et al.} [Event Horizon Telescope],
``First M87 Event Horizon Telescope Results. VII. Polarization of the Ring,''
Astrophys. J. Lett. \textbf{910}, no.1, L12 (2021)
doi:10.3847/2041-8213/abe71d
[arXiv:2105.01169 [astro-ph.HE]]; 

K.~Akiyama \textit{et al.} [Event Horizon Telescope],
``First Sagittarius A* Event Horizon Telescope Results. VII. Polarization of the Ring,''
Astrophys. J. Lett. \textbf{964}, no.2, L25 (2024)
doi:10.3847/2041-8213/ad2df0; 
\bibitem{Faradaysense}
Ferriere,~K. and West,~J.~L. and Jaffe,~T.~R.,
``The correct sense of Faraday rotation,''
MNRAS \textbf{507}, 4968-4982 (2021) [arXiv:2106.03074v1 [astro-ph.GA]]
doi:10.1093/mnras/stab1641
 
\end{thebibliography}
\end{document}